\documentclass[aps,prb,twocolumn,showpacs]{revtex4}

\usepackage{graphicx}
\DeclareGraphicsExtensions{.png,.jpg,.eps,.pdf}

\usepackage{xcolor}
\usepackage{hyperref}
\usepackage{amsmath}

\begin{document}

\title{ Electrical spin manipulation in graphene nanostructures}

\author{ R. Ortiz$^{1}$, N.  A.  Garc\'ia-Mart\'inez$^{2}$, J. L. Lado$^{2,3}$,
J. Fern\'andez-Rossier  $^{2}$}

\affiliation{
$^1$Departamento de F\'isica Aplicada, Universidad de Alicante, 03690  Spain }
\affiliation{
$^2$QuantaLab, International Iberian Nanotechnology Laboratory (INL),
Av. Mestre Jos\'e Veiga, 4715-330 Braga, Portugal}
\affiliation{
$^3$Institute for Theoretical Physics, ETH Zurich, 8093 Zurich, Switzerland
}

\date{\today}

\begin{abstract}
We propose a  mechanism   to drive   singlet-triplet  spin transitions, electrically,  in  a wide class of graphene nanostructures  that present pairs of in-gap zero modes, localized at opposite sublattices.  Examples are    rectangular nanographenes with short zigzag edges,   armchair ribbon heterojunctions with  topological in-gap states  and   graphene islands with sp$^3$ functionalization.
The interplay between the hybridization of zero modes  and Coulomb repulsion leads to symmetric exchange interaction that favors a singlet ground state.
Application of an off-plane electric field to the graphene nanostructure generates  an additional  Rashba spin-orbit coupling, which results in antisymmetric exchange interaction that mixes $S=0$ and $S=1$  manifolds.
      We show  that modulation in time of either the off-plane electric field or the applied magnetic field  permits to perform electrically driven spin resonance in a system with very long spin relaxation times.
\end{abstract}

\maketitle



Spin $1/2$  systems provide the simplest physical realization of a quantum bit\cite{nielsen2002,ardavan2011}. Unsurprisingly,
localized spins, both electronic\cite{kane1998,loss1998,burkard99} and nuclear\cite{gershenfeld1997},  were early on proposed as physical platforms
 to store and manipulate quantum information  taking advantage   from the enormous know-how in magnetic resonance techniques. In spite of several remarkable experimental breakthroughs, using both phosphorous donors in Silicon\cite{muhonen2014} as well as III-V semiconductor quantum dots\cite{elzerman2004,koppens2005},   the fabrication of   spin based quantum computer  in solid state platforms, going beyond a few quantum bits,  remains a daunting challenge.  One of the main problems is the upper limit for  spin coherence lifetimes $T_2$   due to  hyperfine coupling to the nuclear spins  \cite{khaetskii2002}.

  Strategies to mitigate this problem come from two fronts. First, using materials with a small, or even null, density  of nuclear spins, such as graphene\cite{trauzettel2007}  and carbon nanotube based quantum dots\cite{laird2013} or isotopically pure silicon\cite{steger2012}. Second,   using a different degree of freedom to store quantum information, such as the singlet-triplet $S_z=0$ states that arise  for pairs of exchange coupled spins\cite{petta2005}.  However, this approach requires the use of 2 electron spins per qubit, with the resulting fabrication overhead.

\begin{figure}[h!]
 \centering
  \includegraphics[width=0.5\textwidth]{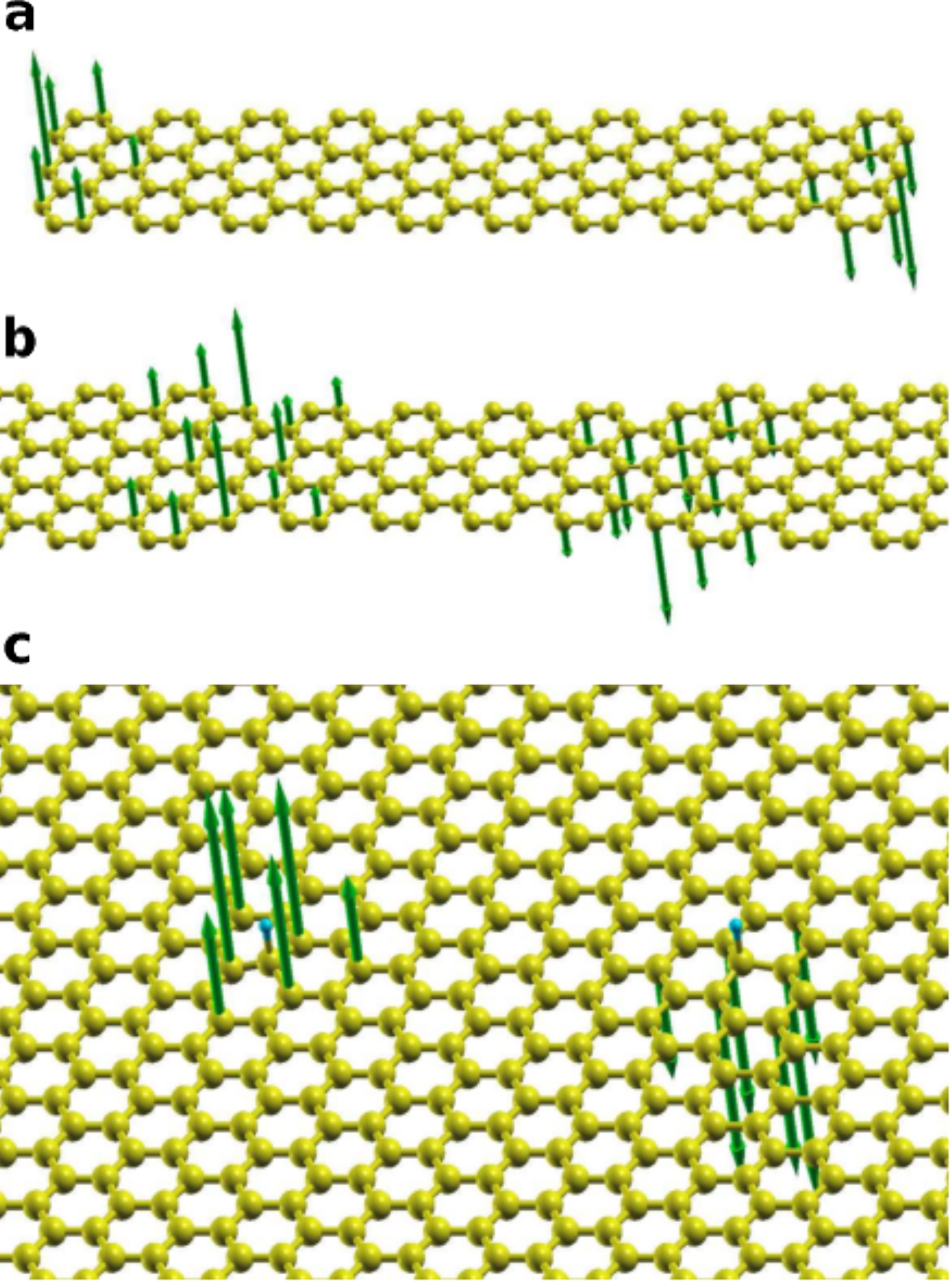}
\caption{ Three types of graphene nanostructures that host pairs of zero modes
localized in opposite sublattices.
$a)$ Rectangular graphene nanoribbons with short zigzag edges that host 1 unpaired electron each.
$b)$  Armchair graphene heterojunctions, hosting 1 zero mode at each interface \cite{cao2017}.
$c)$ Graphene nanoisland with sp$^3$ functionalization (chemisorbed atomic hydrogen).
In the three cases, the  green arrows represent the magnetization calculated with a
mean field Hubbard model. }
\label{Structure}
\label{fig1}
\end{figure}

\begin{figure*}
 \centering
  \includegraphics[width=0.70\textwidth]{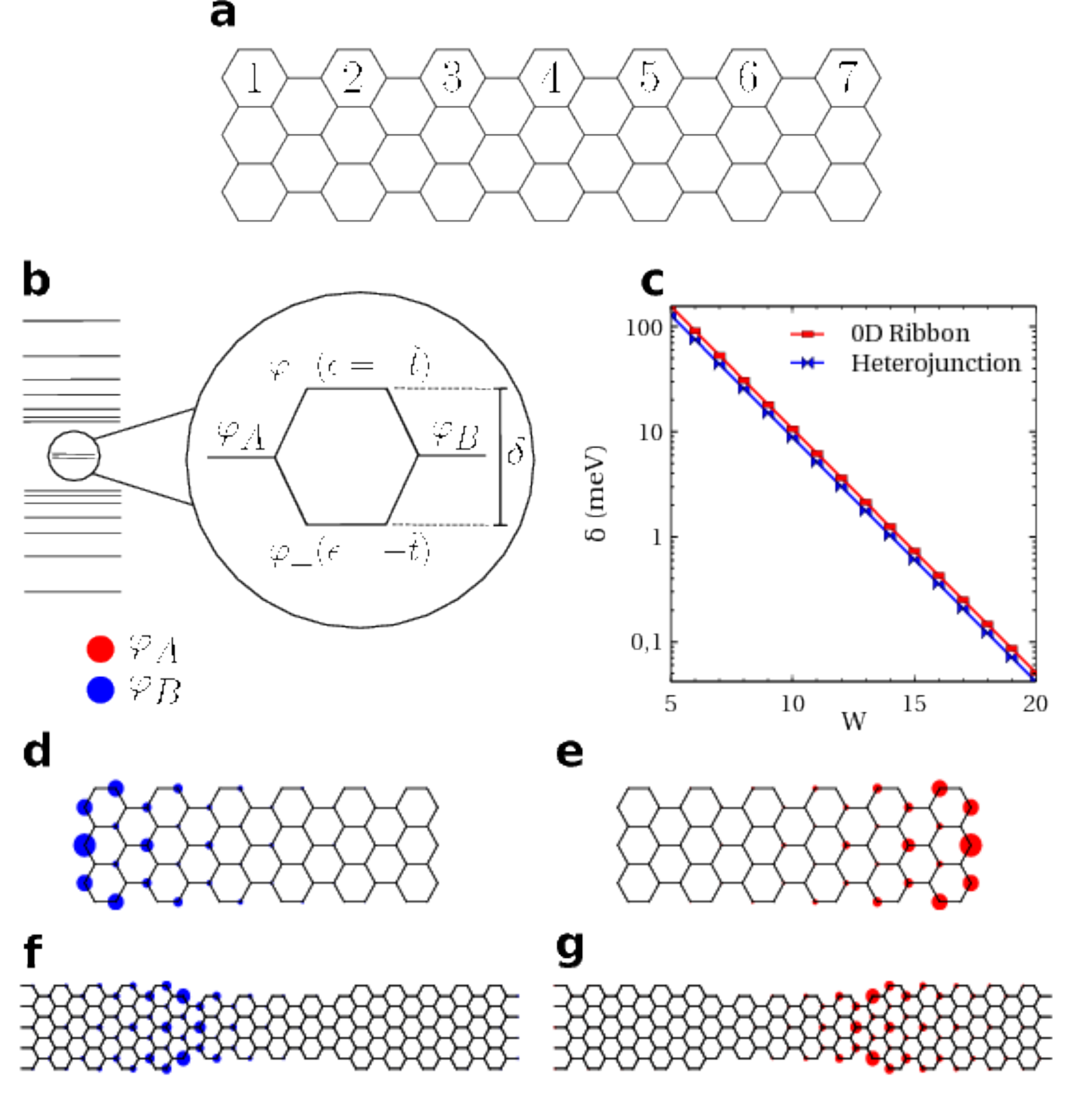}
 \caption{
$a)$  Graphene nanoribbon with $W = 7$.
 $b)$  Sketch of the single-particle energy spectrum for the graphene
nanostructures shown in figure \ref{Structure}.
 A  gap,   separating the doubly occupied states from the empty states,  contains 2 in-gap states,
 $\epsilon_{\pm}$,  split by $\delta=\epsilon_{+}-\epsilon_{-}$.
 $c)$   Dependence of $\delta$ on the  distance between  both for edge and interface states.
 The splitting arises from the hybridization of the zero modes, $\psi_A$ and $\psi_B$,
 for the ribbons (d, e) and heterojunctions (f, g) (see equation  (\ref{zeromodes0})).
 }
\label{figzeromodes}
\label{fig2}
\end{figure*}

Interestingly, a class of graphene nanostructures  that can be synthesized with
bottom-up techniques\cite{wang2016,wang2017} provide naturally,  without the
need of electrical control of the number of carriers,    exchange coupled
unpaired spin electron duets in an environment with a low density of carbon
nuclear spins.
In figure \ref{Structure}  we show three such  graphene nanostructures:  graphene rectangular  ribbons with short zigzag edges (in the following ribbons),   armchair ribbon heterojunctions with topological in-gap states (in the following heterojunctions) and  $sp^3$ functionalized graphene.  These three  systems  form a class with the following common properties:
\begin{enumerate}
\item On account of their finite size,  they have a gapped spectrum,  except
for two single-particle in-gap states, that we label $\psi_{\pm}$, and host two
electrons  (see figure \ref{figzeromodes}(b)).

\item The wave function of these in-gap states turns out to be a linear
combination of two zero mode states that are mostly localized in one
of the
sublattices, labeled  $A$ and $B$ that form the honeycomb lattice (figure
 \ref{figzeromodes}(d, e, f, g)).  We refer to  these zero mode states   as  $\psi_A$
and $\psi_B$.

\item The overlap of $\psi_A$ and $\psi_B$, and thereby the bonding-antibonding
splitting ($\delta\equiv\epsilon_{+}-\epsilon_{-}$) of the single-particle
spectrum,   depends on the geometrical properties of the graphene structure,
and is therefore an important design parameter (figure \ref{figzeromodes}(c)).

\item The electronic ground state is a singlet with $S=0$, the first excited state is a triplet $S=1$ and their energy separation $J$ is proportional to $\delta^2/\tilde{U}$, where $\tilde{U}$ is the  Coulomb energy overhead of adding a second electron in the localized states ($\psi_{A,B}$).
\end{enumerate}

In this work two things are done. First, we provide a quantum  theory, beyond mean field approximation,  for the spin states and the exchange $J$ in this class of graphene nanostructures. Second, we study how the application of an off-plane electric field generates a
Dzyaloshinsky-Moriya (DM) antisymmetric exchange\cite{dzyaloshinsky1958,moriya1960} that could be
used
to  enable spin-transitions between the ground state singlet and the states in the triplet.   Importantly, these  transitions are strictly forbidden, in  the absence of DM interaction, in  conventional electron-paramagnetic resonance experiments, where both spins interact with a dc field $B_0$ and a perpendicular $ac$ field $B_{ac}$ and only transitions that conserve $S$ may be induced.  Therefore, our results pave the way towards electrically driven spin resonance in graphene nanostructures, complementing recent experiments on electrically {\em detected} spin resonance in graphene\cite{mani2012,lyon2017}.

Graphene zero modes  with a wave function localized in a single sublattice were predicted to occur in
zigzag graphene  edges\cite{nakada1996,fujita1996} and around carbon atoms with
sp$^3$ functionalization
\cite{wehling2007,yazyev2007,pereira2008,Palacios08}. Their   direct
experimental observation, by means of scanning tunneling microscopy, has
been reported both   for the edge states of  rectangular nanographenes with short zigzag
edges  \cite{wang2016} as well as for  individual and for pairs of chemisorbed
hydrogen atoms in graphene\cite{ugeda2010,gonzalez2016}.  These sub-lattice
polarized zero modes are expected to host unpaired spin electrons,  giving rise
to the formation of local moments
\cite{fujita1996,son2006a,son2006b,kumazaki2007,JFR2007,JFR2008,Palacios08,Yazyev2010,Lado2014prl,Garcia2017}.
 Sublattice polarized zero modes have recently been predicted \cite{cao2017} to
exist as in-gap topological states at the interface of certain graphene ribbons
with armchair edges, shown in figure \ref{Structure}(b).
Recent progress in  fabrication of graphene ribbon
heterojunctions\cite{ruffieux2016,wang2017} shows that fabrication of this type
of structure is not out of reach of state of the art in nanographene synthesis.

The exploration with STM of some of the  graphene nanostructures studied here
has been demonstrated\cite{gonzalez2016,wang2016,wang2017}.  With this
approach,  the application of  an  off-plane electric field significantly larger than in
conventional field effect transistor geometries is possible. On the other hand,
STM can be used  to carry out electrically driven spin paramagnetic resonance
of individual atoms \cite{Baumann2015,choi2017,Natterer2017} and  coupled spin
$1/2$ atoms \cite{Yang2017}. Therefore,  the electrical manipulation of
localized spin states in graphene seems within reach with state of the art
surface scanning probes.

{\em Modeling}.
We model the single particle states of the  graphene nanostructures both with  the standard one-orbital tight-binding model,  with first neighbor hopping $t=2.7$ eV.
%
 Electron-electron interactions are treated with the Hubbard model, both at the
mean field approximation, including all the single particle states,  or exactly
for the subspace of 2 electrons and 2 orbitals that controls the spin
properties of the studied systems. In the case of graphene nanostructures, it
is well known that mean field Hubbard model calculations  and density
functional calculations  give very similar results\cite{JFR2007,ortiz2016}.
The spin orbit coupling effect considered in the following
will be of Rashba type,\cite{Kane2005,min2006,konschuh2010}
that can be externally modulated with an electric field.

{\em  The non-interacting spectrum.}
  A scheme of the single-particle spectrum characteristic of the gapped graphene with 2 in-gap states
 is shown in figure
\ref{fig2}(b).  The energies and wave-functions of the in-gap states are  denoted by
$\epsilon_\pm$ and $\psi_\pm$ respectively.
  It is always possible\cite{soriano2012} to write  down the  wave function  of a couple of conjugate states, with single-particle energy $E$ and $-E$,  in terms of the same sublattice polarized states $\psi_A$ and $\psi_B$.  Therefore, we write
  \begin{eqnarray}
\psi_A(i) \equiv \frac{1}{\sqrt{2}} \left( \psi_+(i) + \psi_-(i)\right) \nonumber\\
\psi_B(i)\equiv  \frac{1}{\sqrt{2}} \left( \psi_+(i) - \psi_-(i)\right)
\label{zeromodes}
\end{eqnarray}
 In the case of the in-gap states, the  peculiar property of the resulting  $\psi_A$ and $\psi_B$ is that they are {\em spatially separated}.   As a result, the resulting splitting that arises from the hybridization of the zero modes,
 \begin{equation}
 \delta =2 \langle \psi_A|{\cal H}_0|\psi_B\rangle \equiv  2 \tilde{t}
 \label{zeromodes0}
 \end{equation}
 turns out to be  small.  In figure \ref{figzeromodes}(c) we plot $\delta$ for
different nanographenes  as  a function of the  spatial separation between the
zero modes.   It is apparent  and well known\cite{nakada1996} that this quantity decays
exponentially with $W$.    In the limit where $W$ is very large (see
figure \ref{figzeromodes}(c)), $\delta$ vanishes, and the energy of the in-gap
states goes to $E=0$, showing that these sublattice polarized states are
zero modes.\cite{nakada1996}

 Our next task is to demonstrate that in-gap states in these structures hold local moments.
  This  has been established, using either DFT and/or  mean field Hubbard model
calculations,  in the case of  infinitely   long  graphene ribbons  with zigzag
edges\cite{fujita1996,JFR2008}, as well as the small nanoribbons considered
here\cite{ruffieux2016,wang2017},  and also for hydrogenated
graphene\cite{PhysRevB.75.125408,PhysRevB.81.165409,gonzalez2016,Garcia2017}. To the best of
our knowledge, the emergence of local moments in  the case of un-doped
topological junctions has not been explored yet.  We therefore carry out a mean
field Hubbard model calculation (see supplementary material for details)
  to address the emergence of local moments associated to  the topological
in-gap states and, for comparison, the well understood case of graphene
nanoribbons.  For the topological in-gap states, we consider a structure with
periodic boundary conditions and two interfaces, that accommodate one in-gap
state each.
  For $U=t=2.7.eV$, we find  broken symmetry solutions with a finite local
magnetization, $M(i)=\langle S_z(i)\rangle$  that is mostly located in the
region where  either $\psi_A$ or  $\psi_B$ are non-zero, for all structures
except those where $\delta$ is large ({\it i.e.}, those where $\psi_A$ and
$\psi_B$ are strongly hybridized).   This applies both for heterojunctions and
nanoribbons. In the mean field approximation, the transition between
non-magnetic and broken symmetry transitions is abrupt.   The mean field broken
symmetry solutions have lower energy for antiferromagnetic  (AF) correlations
between spins in opposite sublattice, that result in a total zero magnetic
moment $\sum_i M(i)=0$  (see figure \ref{Structure}(b)).      Solutions with a
net magnetic moment and ferromagnetic (FM) correlations between opposite
sublattices have higher energy and  $\sum_i M(i)=1$,  as expected for
as expected for a $S=1$ configuration in
two antiferromagnetically coupled $S=1/2$.

\begin{figure}[h!]
 \centering
  \includegraphics[width=0.5\textwidth]{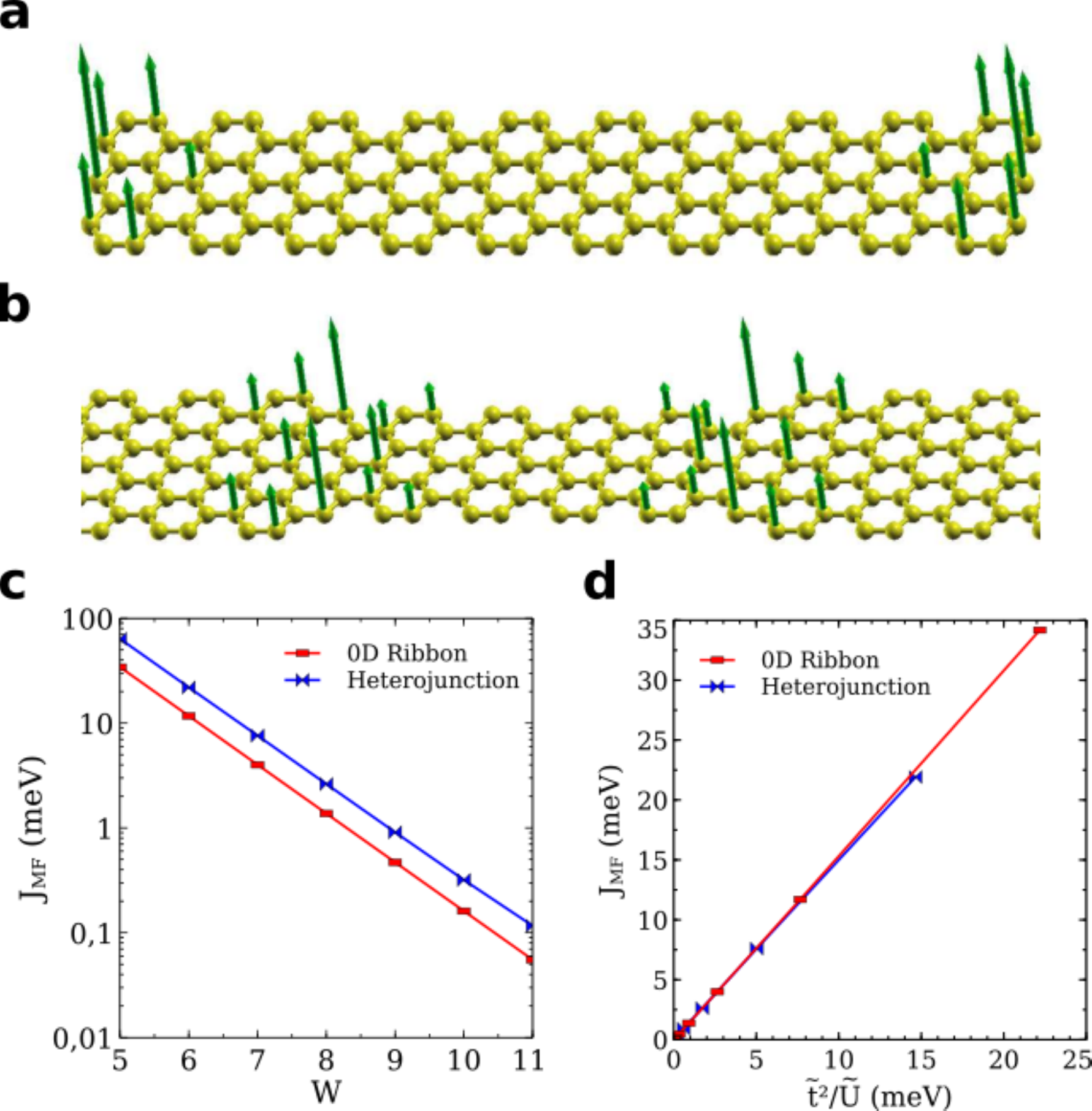}
\caption{$a)$, $b)$ Magnetization in the ferromagnetically aligned (FM) configuration , as calculated within the mean field approximation for graphene ribbon.
$c)$ Dependence of the exchange energy, calculated within the mean field Hubbard model, $J=E_{FM}-E_{AF}$ on the dimensions of the graphene nanostructure and $d)$ scaling of $J$ with $\frac{\tilde{t}^2}{\tilde{U}}$, demonstrating kinetic exchange.  }
\label{Fig3}
\end{figure}

We study the   exchange energy as the  difference between FM and AF solutions $  J_{MF}= E_{FM}-E_{AF}$
 for several different nanographenes, both for the edge and interface states.   We
find that, for the same value of $W$,  the exchange is larger for ribbons than
heterojunctions.  This ultimately arises from the larger hybridization of the edge
zero modes,  compared with the  topological interface zero modes (see figure \ref{fig2}(c)).
  We show in
figure \ref{Fig3}  that $J_{MF}$ can be as large as 40 meV for graphene
ribbons, and be made as small as necessary by increasing the distance $W$
between the zero modes.
 Importantly,  as  we show in figure  \ref{Fig3}(b),  we find that, both for
ribbons and heterojunctions, exchange energy scales as
 \begin{equation}
J_{MF}  \propto \frac{\tilde{t}^2}{\tilde{U}}
  \label{KE1} \end{equation}
  where
    \begin{equation}
\tilde{U}=U \sum_i |\psi_A(i)|^4 = U\sum_i |\psi_B(i)|^4= U\eta
\end{equation}
is the average addition energy for these states, as computed in the Hubbard model (see supplementary material) and $\eta$ is the inverse participation ratio of the zero mode states.  This scaling provides a strong indication that the mechanism  of antiferromagnetic interaction is kinetic exchange\cite{Anderson1959,moriya1960}, that arises naturally for half-filled Hubbard dimers.
Our calculations show that,  for a given type of structures (ribbon or
heterojunction), the inverse participation ratio $\eta$ is quite independent of
$W$. Thus, for the zigzag edge zero  modes we find $\eta\approx0.1$  and for the
topological in-gap states we find $\eta\approx0.035$. The smaller $\eta$ for
the heterojunction states can be anticipated, as they can spread at both sides
of the junction,  in contrast with the edge states.

All these results,  most notably the scaling of equation \ref{KE1}, strongly suggest
that magnetic correlations are governed by the two electrons that occupy the
two in-gap states.  This is also the case  for graphene ribbons with
infinitely long zigzag edges \cite{JFR2008}.  In order to go beyond the mean
field picture and to be able to  describe  local moments in these nanographenes
with a full quantum theory without  breaking symmetry,  we
  restrict the Hilbert space  to the configurations of 2 electrons in the two  zero modes.  To do so,
we represent the Hubbard interaction in the one body basis defined by the states $\psi_A$ and $\psi_B$. The Hamiltonian so obtained is a two site Hubbard model with renormalized hopping and on-site energy:
\begin{equation}
{\cal H}_{\rm eff} = \tilde{t} \sum_{\sigma}\left( a^{\dagger}_{\sigma} b_{\sigma} +  b^{\dagger}_{\sigma} a_{\sigma} \right)
+  \tilde{U} \left( n_{A\uparrow}  n_{A\downarrow}  + n_{B\uparrow}  n_{B\downarrow}   \right)
\label{H2}
\end{equation}
where $a^{\dagger}_{\sigma}=\sum_i \psi_A(i) c^{\dagger}_{i\sigma}$ and $b^{\dagger}_{\sigma}=\sum_i \psi_B(i) c^{\dagger}_{i\sigma}$ are the
operators that create an electron in the zero modes $\psi_A$ and $\psi_B$  with spin $\sigma$, respectively. In turn,  $n_{A,\sigma}= a^{\dagger}_{\sigma} a_{\sigma}$ is the number operator for the $\psi_A$ state with spin $\sigma$.  In addition, we consider the Zeeman coupling to a magnetic field,
\begin{equation}
{\cal H}_{\rm Z}=g\mu_B \sum_{\sigma,\sigma'} \vec{B}\cdot\vec{S}_{\sigma,\sigma'}
\left(a^{\dagger}_{\sigma}a_{\sigma'} +b^{\dagger}_{\sigma}b_{\sigma'}
\right)
\label{Zeeman}
\end{equation}
where $\vec{S}_{\sigma,\sigma'}$ are the $S=1/2$ spin matrices,  $g=2$ is the gyromagnetic factor and $\mu_B=57\mu\, eV\, T^{-1}$ is the Bohr magneton.

\begin{figure*}
 \centering
 \includegraphics[width=0.9\textwidth]{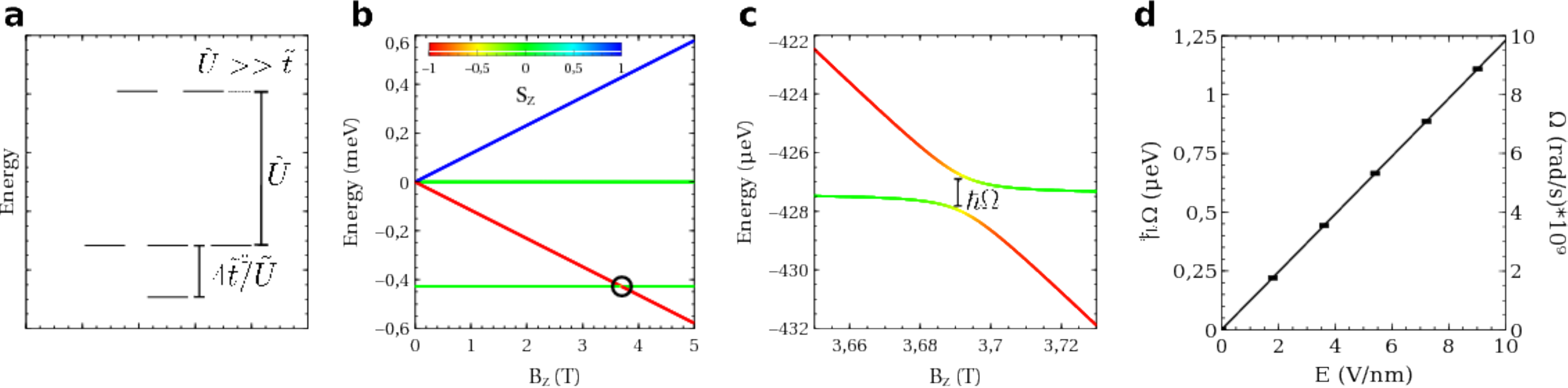}
\caption{
$a)$ Scheme of energy levels for 2 site Hubbard model with 2 electrons.
$b)$ Evolution of 4 lowest energy eigenstates of equation (\ref{H2}), including the Zeeman (equation  (\ref{Zeeman}))
for the ribbon with $W=10$,  $U=t$ and Rashba,  for $E=10\ V/nm$.  The effect of the Rashba interaction is only apparent in the anticrossing of the $S_z=-1$ and $S=0$ states,  shown in the inset.
$c)$ Zoom of the anticrossing.
$d)$  Magnitude of the  singlet-triplet anticrossing energy $\hbar\Omega$ as a function of the electric field.}
\label{Fig4}
\end{figure*}

  Hamiltonian (\ref{H2}) is a two-site Hubbard model, where the sites
correspond to the zero mode states $\psi_{A,B}$, shown in figure
\ref{figzeromodes}(b, c, d, e).  This model  can be solved
analytically\cite{jafari2008} or by a straight-forward numerical
diagonalization (see supplementary material).  For the relevant case of 2 electrons,  the
dimension of the Hilbert space is 6 and the ground state is always a singlet.
We are interested in
   the limit  $\tilde t <<\tilde U$.  A cartoon of the spectrum is shown in
figure \ref{Fig4}(a).  In that case the excited state manifold is a triplet,
way below two excited singlets that describe  states with double occupation of
the zero modes.

      Unlike the mean field solution,  the exact solution of Hamiltonian (\ref{H2}) has  no abrupt change of behavior from non-magnetic to magnetic solutions.
However, depending on the ratio $\frac{\tilde t}{\tilde U}$, the physical properties of the system are very different.  
  %
  This is quantified  by the weight  on the ground state wave function of  the
states where 2 electrons occupy one zero mode, denoted by $P_2$.  For $U=0$ the
ground state is a trivial singlet, formed by two electrons in the lowest energy
in-gap state and $P_2=0.5$.   For very small $\hat t/\hat U$,  $P_2$ goes to
zero.  For a fixed value of $t$ and $U$,  the effective hopping $\tilde{t}$ is
controlled by the dimensions of the nanographene structure. Thus,  in figure
\ref{P2vsthings}(a),  we show $P_2$ for a nanoribbon,  assuming $U=t$, as a
function of the ribbon width $W$.  We see that for $W>9$,  the weight of the
double occupancy configurations  is smaller than 5 percent of the state, and
the charge fluctuations are effectively frozen.    In that limit,
  it is well known\cite{Anderson1959,moriya1960} that the four lowest
levels in the model of equation (\ref{H2})  can be mapped into
 the Heisenberg Hamiltonian:
\begin{equation}
{\cal H}_{\rm Heis} = J_{H}\vec{S}_A\cdot\vec{S}_B
\label{HEIS}
\end{equation}
where $\vec{S}_{A,B}$ are the spin $\frac{1}{2}$ operators describing the
electronic spins localized in states $\psi_A$ and $\psi_B$, respectively  and
$J_{H}\simeq \frac{4\tilde{t}^2}{\tilde{U}}$. The Hamiltonian of equation (\ref{HEIS})
has a ground state singlet ($S=0$)and an excited state triplet with $S=1$,
separated in energy by $\Delta=E(S=1)-E(S=0)=J_H$ (see figure \ref{Fig4}(b)).
heterojunctions.
 Effectively, the upper limit to  $J_H$ is marked by the crossover to the  un-correlated regime, where double occupancy $P_2$ is not negligible.    On the other side,   $J$ can be made exponentially small when the distance between the two zero modes is increased.

  We now consider the effect of spin-orbit interactions induced by an off-plane electric field, $\vec{E}$,  on the spin dynamics of these 4 states.
  These can be described with  a  Rashba spin-orbit coupling\cite{Kane2005,min2006,konschuh2010},
   \begin{equation}
 {\cal H}_{\rm R}=  i t_R \sum_{\sigma,\sigma',\langle i,j\rangle} \vec{E}\cdot\left(\vec{d}_{i,j}\times\vec{\sigma}_{\sigma,\sigma'}\right)
 c^{\dagger}_{i\sigma}c_{j\sigma'}
 \label{rashba}
 \end{equation}
 where $\langle i,j\rangle$ labels first neighbors
and $\vec d_{ij}$
in the vector linking them.
$\sigma=\pm $ labels the
eigenstates of the spin matrix $S_z=\frac{1}{2}\sigma_z$,
$\vec{\sigma}_{\sigma,\sigma'}$ are the Pauli matrices (with eigenvalues $\pm
1$),  and the $c$ and $c^{\dagger}$ are second quantization fermionic
operators. The extrinsic spin-orbit coupling constant
  $t_R$  is zero
  unless an off-plane electric field is applied $E_0\hat{z}$ to break mirror symmetry  \cite{min2006}:
 \begin{equation}
t_R  =  \frac{eE z_0 }{9V_{sp\sigma}}\xi
\label{trE}
 \end{equation}
 where $e$ is the electron charge, $E z_0$ is the voltage drop across
atomically thin graphene\cite{min2006}, $\xi=6 meV$ is the spin orbit coupling
of carbon and $V_{sp\sigma}$ is the hybridization between $p$ and $s$
orbitals.\footnote{In the work of Min et al, the effective Hamiltonian at the
Dirac point is derived, and the Rashba coupling, denoted by  $\lambda_R$ is
used. When that effective Hamiltonian at the Dirac point is derived from the
tight-binding expression (equation \ref{rashba}),    it is found that $\lambda=-3
t_R$.}

 For an electric field
$E=50 {\rm Volt}/300 nm$, standard for graphene field effect
transistors\cite{novoselov2004}, we have $t_R\simeq 3.7\mu$eV.
 \footnote{ From equation (\ref{rashba}) we can derive the effective Hamiltonian at the Dirac points of 2D graphene}
Importantly,  with an STM tip it is possible to apply a few volts at 1 nm, so that  $t_R=100 \mu$eV could be reached.  

 The Rashba spin-orbit Hamiltonian adds an spin-flip
   hopping in the 2-site model (\ref{H2}):
\begin{equation}
{\cal V}_{\rm R} =  \sum_{\sigma}\left( \tilde{t}_R(\sigma) a^{\dagger}_{\sigma} b_{\overline{\sigma}} + \tilde{t}_R (\sigma)^* b^{\dagger}_{\overline{\sigma}} a_{\sigma} \right)
\label{VR}
\end{equation}
where $\overline{\sigma}=-\sigma$ and
\begin{equation}
\tilde{t}_R(\sigma)=\sigma \langle \psi_{A\sigma} |{\cal V}_{\rm R} |\psi_{B\overline{\sigma}}\rangle\equiv \sigma \tilde{t}_R
\label{tR}
\end{equation}
For the graphene nanostructures considered here,  we find that $\tilde{t}_R$ is real.
Unexpectedly, we find that $\frac{\tilde{t}_R}{\tilde{t}}$ is always more than 5 times larger than $\frac{t_R}{t}$. The origin of the enhancement of the Rashba interaction in  graphene nanostructures has to do with a constructive interference between the modulation of the sign of the in-gap zero modes states and the angle-dependence sign of the Rashba hopping.

The addition of this spin-flip hopping to the Hubbard model results, in
 the strong coupling limit $\tilde{U}>>\tilde{t}$, in  two types of additional
terms to the effective spin Hamiltonian\cite{moriya1960,goth2014}:

\begin{eqnarray}
{\cal V}_{\rm DM}= J_{\rm DM} \left[ \left(S_A^xS_B^z-S_A^zS_B^x\right)  +
 \left(S_A^zS_B^y-S_A^yS_B^z\right) \right]
\label{DM}
\end{eqnarray}
\begin{eqnarray}
&&{\cal V}_{\rm anis} =
J_z  S_A^z S_B^z
\label{anis}
\end{eqnarray}
with $J_{DM}=\frac{8\tilde{t}\tilde{t}_R}{\tilde{U}}$ and $J_z=4\frac{\tilde{t}^2-\tilde{t}_R^2}{\tilde{U}}$.

The first term (equation \ref{DM})  is the widely studied anisotropic exchange postulated  by
Dzyaloshinsky \cite{dzyaloshinsky1958} and derived by Moriya\cite{moriya1960}. It does not conserve $S_z$. The physical origin is transparent: exchange arises from the virtual hopping  of one electron between states $\psi_A$ and $\psi_B$.  This hopping occurs through a spin conserving channel, with amplitude $\tilde{t}$ and through a spin-flip channel
$\tilde{t}_R$.  Thus, two hoppings through the same channel, either spin conserving or spin flip,  preserve the spin of the electron. In contrast, the crossed term, by which only one hopping preserves the spin, results in an effective interaction that does not conserve $S_z$.  This is the DM interaction, which is the dominant addition coming from the Rashba perturbation,   given that $\tilde{t}>>\tilde{t}_R$.

The  DM interaction  scales with the kinetic exchange as  $J_{\rm DM}= \frac{\tilde{t}_R}{\tilde{t}}J$. Thus, $J$ is in the range of meV, so that $J_{\rm DM}$ in this system is, at most, in the $\mu$eV.  Whereas this is a small energy scale, it has a qualitatively important consequence: it   permits otherwise forbidden transitions between singlet and triplet manifolds.    This is shown in
figure \ref{Fig4}(b), where  we plot the spectrum of the 2-site Hubbard model,
as a function of the off-plane magnetic field $B$,  for a ribbon with $W=10$,
chosen so that for a  moderate magnetic field  the  Zeeman splitting of the
triplet manifold  offsets the singlet-triplet splitting $J$.  The calculation
is done including the effect of the Rashba interaction. The effect of the small
Rashba interaction  is only apparent when the $S_z=-1$ triplet state gets close
in energy to the $S=0$ ground state. In the absence of Rashba interaction,
these two spectral lines would cross each other.

  We have verified that dipolar interactions (see supplementary material \ref{dipo})
are small  (for the $W=10$ nanoribbon,   $10^{-2}\mu$eV ).  Importantly,   they produce an anisotropic symmetric exchange that  does  not couple $S=0$ with the $S_z=\pm 1$ states.  In addition,  dipole interaction can not be   modulated electrically in this class of systems.

The  states $S_z=-1$ from the triplet and the $S=0$ define a two level system with Hamiltonian:
\begin{equation}
{\cal H}_{\rm TLS} =\frac{1}{2} \hbar\omega_0 \left(\tau_z+1\right) + \frac{\hbar}{2} \Omega \tau_x
\label{TLS}
\end{equation}
where  $\tau_z$ and $\tau_x$ are  the $S=1/2$ Pauli matrices (with eigenvalues $\pm 1$),
$\hbar \omega_0= 
J-g\mu_B B$
  is the splitting of the two levels when the electric field is zero, and
\begin{equation}
\hbar\Omega\propto \frac{\tilde{t}}{\tilde{U}} \tilde{t}_R
\label{Omega}
\end{equation}
is the Rabi coupling. As expected from equations (\ref{trE}, \ref{VR}, \ref{tR} and \ref{Omega}) , we find that $\hbar\Omega$  scales linearly with the electric field (figure \ref{Fig4}(d)). It must be noted that our TLS is different from  the case of singlet-triplet qubits where both states have $S_z=0$. As a result, the energy difference can be tuned with a magnetic field,  but this  also removes the protection against fluctuations of the magnitude of the external magnetic field that makes singlet-triplet qubits convenient\cite{koppens2005}.

The energy scale $\hbar \Omega$ defines a Rabi coupling between the spin split levels.
In order to asses its magnitude, we first compare it with the Rabi coupling  achieved by pumping a  spin $S=1/2$ system with the ac magnetic field of a microwave.
 The magnetic field of a microwave generated in pulsed state of the art ESR setup is, at most, $B_{ac}=4$mT, leading to  a Rabi splitting
 of  $g \mu_B B_{ac} \simeq 0.4 \mu$eV.  Thus,  electrical driving can overcome
conventional microwave coupling, showing that it can be used to efficiently
drive singlet-triplet spin transitions in graphene nanostructures.

In order to assess the strength of the system response to the electrically driven spin resonance, it is important to compare the Rabi coupling, that drives the TLS out of equilibrium, with the spin relaxation $T_1$  and decoherence  $T_2$ times.  For instance,
   the steady state solution of the Bloch equation  for a TLS driven with a resonant $ac$  Rabi coupling  is fully determined by the dimensionless constant $x=\Omega^2 T_1 T_2$ (see supplementary material). Both $T_1$ and $T_2$   depend a lot on whether the nanographenes are deposited on top of a conductor or an insulator.  In the former case, exchange interaction with the electrons in the conductor will be the dominant spin relaxation and decoherence mechanism\cite{Delgado2017}.

We now provide a {\em rough estimate} of the contribution to  $T_2$    coming from an intrinsic mechanism, namely,   the hyperfine coupling with the  nuclear spins of the hydrogen atoms that passivate the carbon atoms.
Given that the natural abundance of spinless $^{12}C$ is 99 percent,  hyperfine interaction with carbon is less important.  In addition,  isotopically pure graphene could be used and get rid of $^{13}C$ completely.    In principle, hyperfine interaction  between the graphene unpaired electronic spins and the edge hydrogens has two components, the contact Fermi interaction and the dipole-dipole interaction.  The former is stronger, in general, and depends on the probability for the electrons in the zero mode states to visit the hydrogen $1s$ orbital.  It can be seen right away that hybridization of the $p_z$ orbitals of carbon with the $1s$ orbital of hydrogen is zero when these atoms lie in the same plane. Therefore,  Fermi contact interaction with edge hydrogen atoms vanishes altogether and we are left with the dipolar coupling.

The electronic spins will  undergo dephasing due to the stochastic addition of
the magnetic field created by the nuclear magnetic moments. In order to
estimate this effect,   we treat  the nuclear moments as   classical
independent random variables $\vec{m}_N$.   The average nuclear magnetic
field is zero, but the standard deviation ${\cal B}_{z}^2$ is not.
We assume that the nuclear spins undergo a stochastic motion with a white noise spectrum with correlation time $\tau$. Under these assumptions, the $T_2$ dephasing time  for the electronic transitions due to their hyperfine interaction with the edge hydrogen atoms is\cite{slichter2013,Delgado2017}
$T_2^{-1}=  \left(\frac{g\mu_B  {\cal B}_{z}  }{\hbar}\right)^2\tau$. This
equation is valid as long as $\tau$ is the shortest time-scale in the
problem\cite{slichter2013,Delgado2017}. In particular, $\tau<<\omega_0^{-1}$,
where $\hbar \omega_0$ is the electronic Zeeman splitting.   Therefore,  in its
range of validity,  the upper limit for the decoherence rate is given by  $
T_2^{-1} < \left(\frac{g\mu_B  {\cal B}_{z}  }{\hbar}\right)\frac{{\cal
B}_z}{B}$. In the supp. mat. we have obtained ${\cal B}_z\simeq 1 mT$. This
small field produces a electronic Zeeman splitting of 120 $neV$.  The resulting
estimate for the decoherence rate is
$T_2>0.5 $ms.   Using  $T_1>T_2$ we can obtain a lower limit for   $x=\Omega^2 T_1 T_2>\Omega^2 T_2^2$.  For $\hbar\Omega= 1\mu eV$,  we obtain $x>>100$.   So,   the intrinsic decoherence mechanism does not pose an obstacle for the  proposed electric manipulation of the spin states of singlet-triplet states in graphene nanostructures.

{\em Discussion and Conclusions}.

We have identified a class of graphene nanostructures that host local spin
moments in the form of pairs of  antiferromagnetically coupled electrons.  We
have presented a full quantum theory for these local moments that goes beyond
the broken symmetry mean-field and DFT based calculations. We have identified a
new mechanism to efficiently drive spin transitions by application of an
off-plane electric field. The mechanism, particularly efficient in graphene
nanostructures, relies on the  electrically driven breakdown of mirror symmetry
that generates  of spin-orbit coupling  in the single-particle wave functions.
In turn, this induces  and antisymmetric Dzyaloshinsky-Moriya  exchange in the
spin Hamiltonian that mixes the $S=0$ ground state with the $S_z=\pm 1$ states
of the triplet. The strength of the Rabi coupling is found to exceed the one
obtained for $S=1/2$ with state of the art conventional spin resonance driven
with microwaves.  Importantly, the proposed mechanism permits to drive
transitions that are forbidden in conventional spin resonance experiments.

The proposed  mechanism  is different from other proposals for electrically
driven spin resonance. Some of them  rely on the modulation of the crystal
field Hamiltonian \cite{Baumann2015,Ardavan2013}. Others, on  the slanting
magnetic\cite{pioro2008} or exchange\cite{Lado2017} field of a nearby magnetic
electrode.   Our findings could be used to manipulate individual pairs of
spins in nanographene structures.  The independent progress both in spin resonance driven by scanning tunneling
microscopes and  in the fabrication of atomically defined  graphene nanostructures with   bottom-up techniques\cite{wang2016,wang2017,Pascual2017,Cardoso2017},
could permit
to explore their potential for spin qubits.



{\em Acknowledgments}

 This work has
been financially supported in part by FEDER funds
We acknowledge financial support by Marie-Curie-ITN
607904-SPINOGRAPH,  FCT, under the project PTDC/FIS-NAN/4662/2014, and
 MINECO-Spain (MAT2016-78625-C2).
  N. Garcia and J. L. Lado  thank
the hospitality of the Departamento de F\'isica Aplicada at
the Universidad de Alicante.

\appendix

\section{Mean field Hubbard model}

The exact solution for the Hubbard model is only possible  in some very
specific instances, such as a 1d chain, by means of Bethe antsaz,  or in small
clusters via numerical diagonalization. For the nanographenes considered
here,
we  make use of the so called mean field
approximation,\cite{fujita1996,JFR2007,PhysRevB.77.195428,JFR2008,Lado2014prl,Garcia2017}
where the exact 4-fermion operator is replaced by
\begin{equation}
 {\cal V}_{MF} = U\sum_i (n_{i\uparrow} \langle n_{i\downarrow}\rangle +
n_{i\downarrow}\langle n_{i\uparrow}\rangle-
\langle n_{i\downarrow}\rangle \langle n_{i\uparrow}\rangle
)
\end{equation}
where $\langle n_{i\sigma}\rangle$ stands for the average  number operator, evaluated with the eigenstates of the mean field Hamiltonian obtained from the sum of ${\cal V}_{MF}$ and the single-particle part.  Of course, this defines a self-consistent problem, that is solved by numerical iteration.  Depending on the atomic structure of the nanographene, and the ratio $U/t$, the mean field self-consistent solutions can describe broken symmetry solutions with local moments, or non-magnetic solutions.

\section{ Exact solution of 2 site Hubbard model }
\label{2site}

The Hilbert space for the 2 site Hubbard model with 2 electrons (half filling) has a dimension of 6, spanned by the basis set of Fock states in the site representation
 $(2,0)$, $(0,2)$,  $(\uparrow,\uparrow)$, $(\downarrow,\downarrow)$, $(\downarrow,\uparrow)$ and $(\uparrow,\downarrow)$
 with a self-evident notation, so that the first (second) state represents a doubly occupied $A$ ($B$) site,  the third state denotes the two sites with single occupation with a $S_z=+1/2$ each, and so on.  In this basis set, the Hamiltonian matrix is readily calculated, taking into account the sign that arises from
 the definition of the Fock states in terms of the second quantization operator, as:
 \begin{eqnarray}
 {\cal H}= \left(
\begin{array}{cccccc}
 \tilde{U}&0 &-\tilde{t}_R &-\tilde{t}_R &-\tilde{t} & \tilde{t}\\
 0&\tilde{U} &-\tilde{t}_R &-\tilde{t}_R &-\tilde{t} & \tilde{t}\\
 -\tilde{t}_R&-\tilde{t}_R & g \mu_B B_z&0 &0 & 0\\
 -\tilde{t}_R&-\tilde{t}_R &0 &-g \mu_B B_z &0 & 0\\
 -\tilde{t}&-\tilde{t} &0 &0 &0 & 0\\
 \tilde{t}&\tilde{t} &0 &0 &0 &0
\end{array}
 \right)
 \label{2sitesham}
 \end{eqnarray}

\begin{figure}
 \centering
  \includegraphics[width=0.5\textwidth]{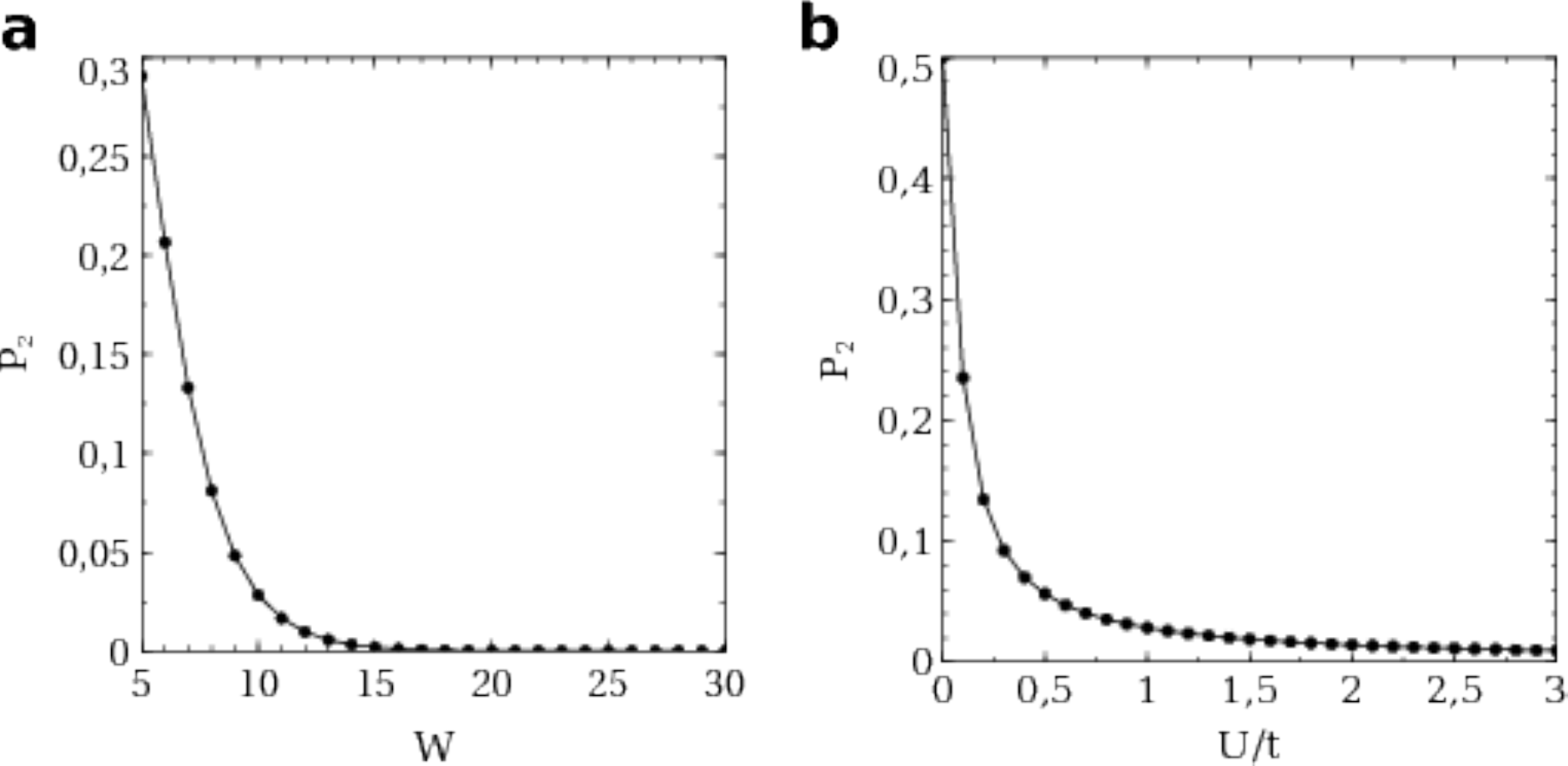}
\caption{$a)$ representation of $P_2$, the weight of the double occupancy states on the ground state wave function,  for graphene ribbons as a function of $W$ (for t = 2.7 eV and $U = t$).  $b)$ $P_2$ as a function of  $U/t$, for the ribbon  with $W=10$.}
\label{P2vsthings}
\end{figure}

For $t_R$ and $B_z=0$, and in the relevant limit with $\tilde{t}<<\tilde{U}$,  the eigenvalues are, in increasing order of energy,  a singlet, a triplet, and two more non-degenerate singlets (see Figure \ref{Fig4}(a)). We define   the weight of  the $(2,0)$ and $(0,2)$ configurations on the ground state singlet,  $P_2= |\langle 20|\Psi_G\rangle|^2  +|\langle 02|\Psi_G\rangle|^2$.  The smaller $P_2$, the better the approximation of the spin model to describe the singlet and triplet states.
The dependence of $P_2$ on $W$  and  $U/t$ is shown in figure \ref{P2vsthings}  for rectangular graphene nanoribbons.  It is apparent that, except for very small for $U=t$ and $W>10$, $P_2$ is below $0.05$.  It is also apparent that there is a smooth crossover from the non-interacting limit, for which $P_2=0.5$,  and the local moment limit for which charge fluctuations are frozen.

\section{Electronic dipolar interaction}
\label{dipo}
 Here we consider the effect of the dipole-dipole coupling between the magnetization cloud of state $\psi_A$ with state $\psi_B$. This leads to an additional term in the spin Hamiltonian:
 \begin{equation}
 {\cal H}_{\rm dip} = \sum_{a,b} D_{ab}S_a(1)S_b(2)
 \label{dip0}
 \end{equation}
 where $a=x,y,z$  and
 \begin{equation}
 D_{ab}=(g\mu_B)^2\frac{\mu_0}{4\pi} \Lambda_{ab}
 \label{dipolartensor}
 \end{equation}
where
 \begin{equation}
\Lambda_{ab}= \sum_{i,i'} |\psi_L(i)|^2|\psi_R(i')|^2
 \frac{\delta_{a,b} -3 n_{a}(ii') n_{b}(ii')}{r_{ii'}^3}
 \end{equation}
where $n_a(ii')$ is the $a$ component of the unit vector
$\vec{n}(ii')=\frac{1}{|\vec{r}_i-\vec{r}_{i'}|}\left(|\vec{r}_i-\vec{r}_{i'}|\right)$. Of course, the carbon positions lie in the plane $z=0$ so that the $n_z$ components are zero.   Thus, we have:
 \begin{equation}
\Lambda_{zz}= \sum_{i,i'}
 \frac{|\psi_L(i)|^2|\psi_R(i')|^2  }{r_{ii'}^3}
 \end{equation}

\begin{figure}
 \centering
  \includegraphics[width=0.45\textwidth]{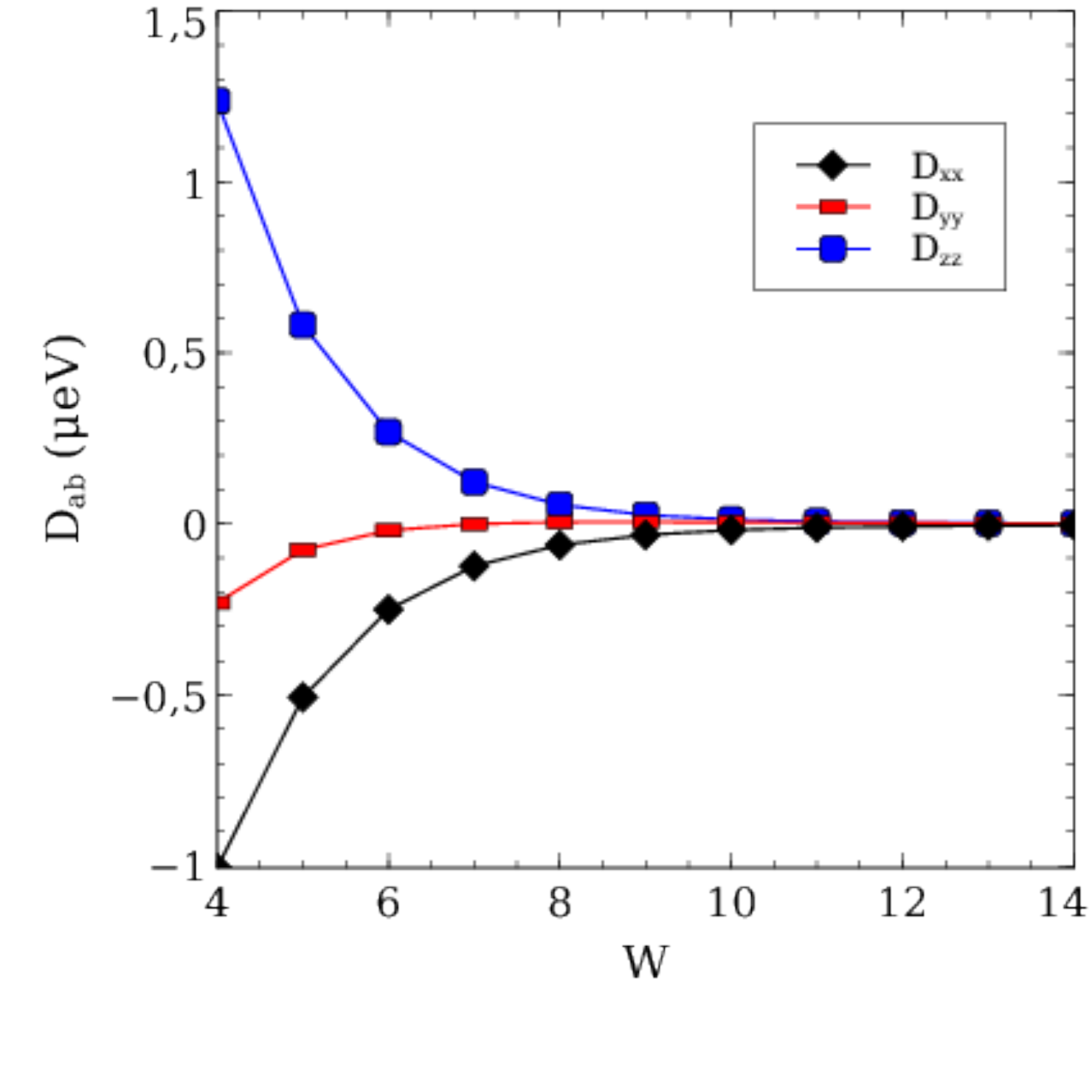}
\caption{Dipolar interaction, as defined in equation (\ref{dipolartensor}), for rectangular graphene nanoribbons, as a function of ribbon size $W$. }
\label{figdipolar}
\end{figure}

Our numerical calculations confirm that  only the diagonal terms of the tensor are finite, as expected from symmetry.
We show them  in figure \ref{figdipolar} for rectangular graphene nanoribbons.  The
elongated shape of ribbons, accounts for the difference between $D_{xx}$ and f $D_{yy}$.  The
resulting  dipolar Hamiltonian can be written as:
\begin{equation}
 {\cal H}_{\rm dip} = - D_{xx}S_x(1)S_x(2)+ D_{zz}S_z(1)S_z(2)
 \label{dip0}
 \end{equation}
 Importantly,  this Hamiltonian does not couple states with different total $S_z$. Therefore, the dipolar interaction does not  couple the two states in the two level system formed by the $S=0$ ground state with the $S_z=-1$ state (equation (\ref{TLS})).  The only effect of the dipolar interaction is to introduce a small anisotropy splitting in the triplet manifold.

\section{Hyperfine  interaction}

The hyperfine interaction is the sum of two dominant
contributions\cite{slichter2013},  Fermi contact interaction and dipolar
coupling. The first is  given by the overlap of the electronic quantum state
with the nuclear species in question.  The Fermi-contact contribution to the
hyperfine interaction of the edge electron, $A$ or $B$,  on a given hydrogen
atom, denoted with the label $N$,  is computed by calculating the weight of the
wave function on the  $s$ orbital of that atom and multiplying the weight to
the hyperfine interaction of atomic hydrogen, 1024 $MHz$.
In order to estimate the contact  interaction we adopt a tight-binding model that permits to compute how the $\pi$ orbitals of graphene hybridize with the $s$ orbital of hydrogen.  This can be done using the
TB model
with 4 orbitals per carbon atom\cite{fratini2013,garcia2015}, and one orbital per hydrogen atom.    Within this  model,  the mid-gap states are, in principle, a linear combination of $p_z$, $p_x$, $p_y$ and $s$ orbitals of the carbon atoms and the $s$ orbital of the edge hydrogen atoms. However, for flat structures with mirror symmetry, the $p_z$ orbitals are odd under reflection, and are thereby perfectly decoupled from all the other states of the basis set, that are even.  As a result, within this  model we find that the Fermi contact contribution to the hyperfine interaction vanishes for the mid-gap states, as well as all the low energy states,  as long as the edge hydrogen atoms remain in the same plane than the nanographene, which is their equilibrium position.


We thus are left with  hyperfine dipolar coupling, whose magnitude we estimate here.  Since we are interested in the decoherence induced by the nuclear spins on the electronic states, we treat the nuclear spins as classical magnetic moments $\vec{m}_N$, whose orientation is completely random.  At any given time they create a magnetic field  at a carbon site $\vec{r}_i$
\begin{equation}
\vec{B}_i[\vec{m}_N]= \frac{\mu_0}{4\pi} \sum_{N} \frac{\vec{m}_N-3\vec{n}_{Ni} \left(\vec{n}_{Ni}\cdot\vec{m}_N\right)}
{|\vec{r}_N-\vec{r}_i|^3}
\end{equation}
where the index $N$ runs over the edge hydrogen atoms and $\vec{n}_{Ni}$ is the unit vector along the direction that joins the nuclear spin $N$ and the carbon site $i$.  We now write down the electronic magnetization density as:
\begin{equation}
\vec{m}_e(i) =\frac{1}{2} \vec{\tau}_{\sigma,\sigma'}\left( |\psi_A(i) |^2 a^{\dagger}_\sigma a_{\sigma'} + |\psi_B(i)|^2
b^{\dagger}_\sigma b_{\sigma'}  \right)
\end{equation}
where  $\vec{\tau}_{\sigma,\sigma'}$ are the spin $1/2$ Pauli matrices with eigenvalues $\pm 1$.
The dipolar  hyperfine interaction reads:
 \begin{equation}
  {\cal V}_N=- \sum_i \vec{m}_e(i)  \cdot \vec{B}_i[\vec{m}_N]
  \end{equation}

  It is now convenient to define the average nuclear magnetic field by the electronic states:
      \begin{equation}
   \vec{\cal B}_{A,B}=  \sum_i |\psi_{A,B}(i)|^2 \vec{B}_i[\vec{m}_N]
   \end{equation}

This permits to write the interaction of the electronic spins in states $A$ and $B$ with the nuclear spins as:
  \begin{equation}
  {\cal V}_N= \sum_{\sigma,\sigma'}
  g \mu_B 
  \left(
  \vec{\cal B}_A\cdot \vec{S}^A_{\sigma,\sigma'}
  +
  \vec{\cal B}_B\cdot \vec{S}^B_{\sigma,\sigma'}
   \right)
    \end{equation}
    where
    \begin{equation}
 \vec{S}^A_{\sigma,\sigma'}=\frac{1}{2} \vec{\tau}_{\sigma,\sigma'}a^{\dagger}_\sigma a_{\sigma'} ,
\;\;\;
 \vec{S}^B_{\sigma,\sigma'}=\frac{1}{2} \vec{\tau}_{\sigma,\sigma'}b^{\dagger}_\sigma b_{\sigma'}
 \end{equation}
 In the strong coupling limit $\tilde{U}>>\tilde{t}$ this results in the addition of the stochastic magnetic field to the Zeeman contribution in equation (\ref{Zeeman}).

 The nuclear field component along the $z$ direction modifies
  the energy of the $S_z=1$ state of the  TLS, and leaves the energy of the $S=0$ unchanged.  Therefore,
  it induces a shift of the  the TLS splitting,
   defined by equation (\ref{TLS}), by an amount
\begin{equation}
 \delta \omega_0= \frac{g\mu_B}{\hbar}  \left( {\cal B}_{z,A}  + {\cal B}_{z,B}\right)
\end{equation}
which is a functional of the nuclear magnetic moments. { For nanoribbons and heterojunctions, the  mirror symmetry of the structures gives  ${\cal B}_{z,A}  = {\cal B}_{z,B}\equiv {\cal B}_{z,B}$}.

We take the orientation of the nuclear moments as random variables with an uniform distribution,  given that even at mK temperatures,  nuclear Zeeman splitting is much smaller than $k_B T$:
\begin{equation}
\langle \vec{m}_N\rangle=0 \,\,\; ,
\langle m_N^a m_{N'}^{a'}\rangle= \delta_{a,a'} \delta_{N,N'} \frac{m_0^2}{3}
\end{equation}
where $m_0$ is the proton magnetic moment.

As a result, its straightforward to see that the average over nuclear moment realizations vanishes, $\langle \vec{\cal B}_{A.B}\rangle=0$.  The  standard deviation of the components, defined as:
\begin{equation}
{\cal B}_{a,A}^2= \frac{\mu_0^2}{(4\pi)^2}\frac{m_0^2}{3}\sum_{i,i',N} \frac{|\psi_A(i)|^2 |\psi_A(i')|^2 }{r_{iN}^3r_{i'N}^3}
\eta_a(N,i,i')
\label{nucldev}
\end{equation}
where
\begin{equation}
\eta_a(N,i,i')\equiv
 1+ 9 n_{Ni}^an_{Ni'}^a \vec{n}_{Ni}\cdot\vec{n}_{Ni'}
   -   3  ((n_{Ni}^a)^2+  (n_{Ni'}^a)^2)
\end{equation}

In the case of the $a=z$ component we have $n^z=0$ for all $N$ and $i$.  We can obtain a quick estimate for the edge states in the graphene nanoribbons  if we  approximate the wave function as equally distributed in 5 edge carbon atoms and only consider their coupling to the first neighbor hydrogen. In that case, we have:
\begin{equation}
{\cal B}_{z}^2\simeq \frac{\mu_0^2}{(4\pi)^2}\frac{m_0^2}{3}\frac{1}{ d_{HC}^6}
\equiv \frac{1}{3} \left(b_0\right)^2
\label{E11}
\end{equation}
where $d_{H,C}\simeq 1.1 \AA$  is the carbon-hydrogen bond length and $b_0\simeq 1mT$ is the magnitude of the magnetic field created by a proton at a distance $d_{HC}$.
From this, we can estimate the associated shift $\hbar \delta \omega_0 \simeq
120 neV$. {Our numerical calculation of  (\ref{nucldev}) yields ${\cal
B}_{z}=0.2$ mT  for a nanoribbon with $W=10$, in line with the estimate of equation
(\ref{E11})  }

\section{Steady State solution of driven two level system}

The steady state solution of the Bloch equation for a two level system driven by   an a.c.   Rabi  monochromatic signal with  frequency $\omega$ is given by\cite{Baumann2015}
   \begin{equation}
P_0-P_1=\delta P_{\rm eq}
\left ( 1 -\frac{\Omega^2T_1T_2}{1+
 (\omega-\omega_0)^2T_2^2+ \Omega T_1 T_2}
\right )
\end{equation}
where $P_0$ and $P_1$ are the non-equilibrium occupation of the ground and excited states in equation (\ref{TLS})
and $\delta P_{\rm eq} \equiv \tanh\left( \frac{\hbar\omega_0}{2 k_B T}\right)$ is the equilibrium population imbalance.
Thus, a relevant figure to assess the merit of  the electrical control of the spin on electrically driven graphene nanostructures
is  $x^2=\Omega^2 T_1 T_2$.   In resonance, we have $P_0-P_1=\delta P_{\rm eq} \frac{1}{1+x^2}$.  Thus, the maximal departure from equilibrium is obtained for very large $x$.

\bibliographystyle{apsrev4-1}
\bibliography{biblio}{}

\end{document}